\documentclass{iaus}
\usepackage{graphicx}
\usepackage{lscape}
\def\la{\mathrel{\hbox{\rlap{\hbox{\lower4pt\hbox{$\sim$}}}\hbox{$<$}}}}
\def\ga{\mathrel{\hbox{\rlap{\hbox{\lower4pt\hbox{$\sim$}}}\hbox{$>$}}}}
\def\kms {\ifmmode{{\rm ~km\,s}^{-1}}\else{~km~s$^{-1}$}\fi}

\def\citet#1{#1}
\def\citep#1{#1}
\def\bp{~~$\circ$\ }

\title[Masers in AGN] 
{Masers in AGN environments}

\author[Greenhill]   
{L.J. Greenhill}

\affiliation{Harvard-Smithsonian Center for Astrophysics, 60
Garden St, Cambridge, MA 02138 USA.
\break email:  greenhill@cfa.harvard.edu\\[\affilskip]
}

\pubyear{2007}
\volume{242}  
\pagerange{xxx -- yyy}
\date{xxx and in revised form yyy}
 \jname{Astrophysical Masers and their
Environments} \editors{J.M. Chapman \& W.A. Baan, eds.}
\begin{document}

\maketitle

\begin{abstract}

Galactic nuclei are well known sources of OH and H$_2$O maser
emission. It appears that intense star formation in ultra-luminous
infrared galaxies drives most OH sources.  In contrast, nuclear
activity  appears to drive most H$_2$O sources.  When H$_2$O
emission originates in accretion disk structures, constrained
geometry and dynamics enable robust interpretation of
spectroscopic and imaging  data.  The principal science includes
study of AGN geometry at parsec and sub-parsec radii and
measurement of geometric distances in the Hubble Flow.  New high
accuracy estimates of the ``Hubble constant, H$_\circ$" obtained
from maser distances may enable new substantively improved
constraints on fundamental cosmological parameters (e.g., dark
energy).

\end{abstract}

\firstsection 

\section{OH and H$_2$O masers in AGN}

Molecular emission is widespread among active galactic nuclei
(e.g., Henkel \etal\ 1990, Garc\'ia-Burillo \etal\ 2007, and
references therein).  However, maser action is comparatively rare,
having been detected only for OH and H$_2$O.

In general, OH maser sources lie in luminous infrared galaxies.
Seven have been resolved on scales $\la 100$\,pc. Most are
associated at least in part with inclined circumnuclear disks,
inferred from position-velocity gradients or limb brightened
structures (see review by Pihlstr\"om 2005b and case studies by
Momjian \etal\ 2006; Pihlstr\"om \etal\ 2001, 2005a; Richards
\etal\ 2005; Rovilos \etal\ 2003; Trotter \etal\ 1997; Yates
\etal\ 2000).  Estimated radii are $10^1$ to $10^2$\,pc and
dynamical masses are  $10^6$ to $10^8$M$_\odot$. However,
connections between these structures and nuclear activity have yet
to be demonstrated.  Observational hurdles include  separating
star forming and AGN processes, and identifying direct  links
between the OH structures and accretion or outflow (e.g., Vignali
\etal\ 2005; Kl\"ockner \& Baan 2005; Baan \& Kl\"ockner 2006).

In contrast, connections between structures traced by H$_2$O
masers and nuclear activity are reasonably well established. Water
maser host galaxies are in nearly every case readily identified as
AGN; luminosity from star formation is secondary.  (A few H$_2$O
masers are seen in star forming  galaxies~-- Henkel \etal\ 2005;
Greenhill 2002a, and references therein~-- and five H$_2$O hosts
exhibit OH emission as well  (Gallimore \etal\ 1996; Tarchi \etal\
2007)~--  but these constitute a small minority.)   Where H$_2$O
emission has been mapped, and the underlying structures have been
resolved, they are typically found to lie a few tenths to a few
parsecs from the central engines,  and to exhibit dynamical
signatures characteristic of accretion and outflow.

In light of greater understanding of the association between
H$_2$O masers and AGN processes, the remainder of this review will
focus on H$_2$O, leaving the complex and rapidly evolving story of
OH to treatment elsewhere in this volume.

\section{History}

Early interpretations of extragalactic H$_2$O maser emission in
nuclei (mistakenly) focused on star formation as driver. It was a
familiar context.  The NGC\,4945 H$_2$O maser was the first
discovered in a nucleus (Table\,1), prior to identification of an
AGN therein (Dos Santos \& L\'epine 1979).  The second nuclear
H$_2$O source was in a well-known Seyfert galaxy, Circinus, but
this too was attributed to star formation (Gardner \& Whiteoak
1982). Claussen \& Lo (1986) delivered the first strong challenge,
with evidence that nuclear H$_2$O maser emission could be compact
on scales of 3-10\,pc and in close proximity to central engines.
In light of the proposed unification paradigm for AGN, Claussen \&
Lo conjectured that the emission arose from warm gas in the now
canonical torus.  Nearly a decade later, studies of NGC\,4258 were
the first to demonstrate that the emission in fact originates from
annuli in the still more compact, relatively thin disks at radii
on the order of 0.1 to 1 pc (Miyoshi \etal\ 1995; Greenhill \etal\
1995).  Two additional contexts for maser emission were identified
later: (1) downstream from central engines in association with
radio jets (Gallimore \etal\ 1996; Claussen \etal\ 1998;
Sawada-Satoh, this meeting) and (2)  wide-angle outflows of AGN
narrow-line regions (Greenhill \etal\ 2003a).

\begin{table}[bh]
  \begin{center}
  \caption{Historical Perspective}
  \begin{tabular}{lllc}\hline
Date $^{(a)}$ & {\hfill Event\hfill}\ & Target & Reference \\
\hline
11/76  & First extragalactic $\lambda 1$cm H$_2$O maser & M\,33
          & 1 \\
09/77  & First H$_2$O maser  in a galactic nucleus
          & NGC\,4945$^{(b)}$ & 2 \\
04/82  & First H$_2$O maser in a known AGN
          & Circinus                    & 3 \\
09/83  & Detection of compact structure with VLA \& VLBI
          & \hspace{-0.2in}NGC1068/4258                 & 4 \\
10/84  &   & NGC\,4258  & 5 \\
          & \hspace{0.5in} {\it -------- First
Lull in Detection of New Masers --------}
          &   \\
01/92  & High-velocity emission discovered
          &NGC\,4258               & 6 \\
03/93  & Narrow-band surveys target optical AGN
          &                                   & 7,8 \\
10/94  & Demonstration that masers can trace accretion disks
          & NGC\,4258            & 9 \\
04/94  &
          &                                   & 10 \\
08/95  & Demonstration that masers can trace radio jets
          & NGC\,1068                & 11 \\
11/95  &
          & NGC\,1052                & 12 \\
06/97  & Demonstration that masers can trace AGN winds
          &                                   &  13 \\
       & \hspace{0.5in} {\it -------- Second Lull in Detection of New Masers --------}
       &   \\
12/01  & Wide-band surveys begin at DSN 70m antennas
          &                                  & 14,15 \\
04/02  & Wide-band surveys begin at GBT 100m antenna
          &                                  &  16\\
12/03  & HST program begins to X-calibrate Cepheid/maser distances
          &         & 17 \\
01/05  & First maser discovered in a quasar ($z\sim 0.7$)
          & J0804+3607         & 18 \\
03/05  & First submm H$_2$O maser discovered in an AGN & NGC\,3079
          & 19 \\
11/06  & Water Maser Cosmology Project begins (NRAO/CfA/MPI)
          &         & 20 \\
\\
\hline
\end{tabular}
\end{center}
$^{(a)}$ {Observing dates associated with events in column 2.} \\
$^{(b)}$ {L\'epine \& Dos Santos (1977) reported an earlier detection possibly toward the nucleus of  NGC\,253, but independent studies failed to confirm it. See review by Greenhill (2002a).} \\
References: (1) Churchwell \etal\ (1977), (2) Dos Santos \& L\'epine (1979),
(3) Gardner \& Whiteoak (1982), (4) Claussen \& Lo (1986),
(5) Claussen \etal\ (1988), (6) Nakai, Inoue, \& Miyoshi (1993),
(7) Braatz \etal\ (1996), (8) Greenhill \etal\ (2002b),
(9) Greenhill \etal\ (1995), (10) Miyoshi et al (1995),
(11) Gallimore \etal\ (1996), (12) Claussen \etal\ (1998), (13) Greenhill \etal\ (2003a),
(14) Greenhill \etal\ (2003b), (15) Kondratko \etal\ (2006a), (16) Braatz \etal\ (2004),
(17) Macri \etal\ (2006), (18) Barvainis \& Antonucci (2005), (19) Humphreys \etal\ (2005), and
(20) Braatz \etal\ (this volume).\\
\end{table}

\section{Surveys}

Water maser sources in nuclei are rare. The historical record of
search programs is full of hypothesized correlations and physical
arguments intended to identify selection criteria that would
enable detection rates $\gg$ a few percent.  Though small samples
have achieved detection rates as large as $\sim 50\%$ (Henkel
\etal\ 2005), no large survey (N$ >10^2$) has exceeded 1-10\%.
The most consistently successful sample is optically identified
AGN, as first pursued in the mid-1990s (e.g., Braatz \etal\ 1996).
Nearly all masers lie in Seyfert 2 objects and Low Ionization
Nuclear Emission Regions (LINERs), though some are known to lie in
ostensibly inactive galaxies and transition objects between
Seyfert\,1 and~2.  None are known to lie in  ``pole-on'' active
systems.  The incidence rate of maser emission among known
Seyfert-2 objects and LINERs is on the order of 20\% for
$cz<5000$\kms, within which past surveys are believed to be
reasonably complete.  The incidence rate drops with increasing
velocity, probably because of incompleteness and sensitivity
limitations (Braatz \etal\ 2004; Kondratko \etal\ 2006a).  Today,
{\it there are $\sim 90$ known masers in AGN (published and
unpublished) with  $cz<20000$\kms} (Figure~1), plus the maser seen
toward quasar J0804+3607 at $z\sim 0.66$ (Barvainis \& Antonucci
2005).  For the time being, the internal structure and physical
nature of this maser is not known.

\begin{figure}[th]
\begin{center}
\includegraphics[scale=0.4,angle=90]{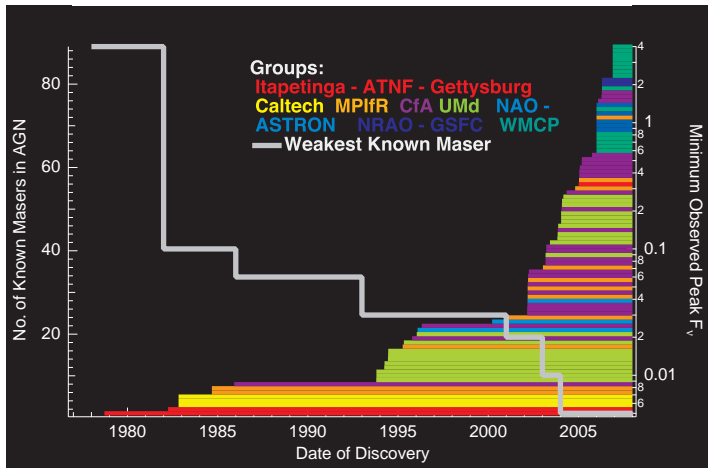}
\caption{Growth in the number of H$_2$O masers believed to lie in
AGN ({\it horizontal bars}). Color coding associates discoveries
with lead institutions.  The evolution in the peak flux density of
the weakest known maser ({\it down-sloping curve}) marks gradual
improvement in observing systems.}
\end{center}
\end{figure}

Because of low detection rates, maser surveys often test stamina,
whether comprising shallow integrations on a large number of
objects or deep integrations on a small number.  The largest
surveys have been made with the Green Bank Telescope, NASA Deep
Space Network, and Effelsberg 100-m (e.g., Braatz \etal\ 2004;
Greenhill \etal\ 2003b; Braatz \etal\ 1996), with important work
on smaller samples also completed at Nobeyama (e.g., Nakai \etal\
1995).   As of now, $> 2000$ galaxies outside the Local Group had
been targeted since the early work by Henkel \etal\ (1984) and
Claussen \etal\ (1984).\footnote{See
www.cfa.harvard.edu/wmcp/surveys/survey.html for cumulative lists
of nondetections.}  Another $\sim$660 ostensibly normal galaxies
have been observed as well (Braatz, Greenhill, unpublished).

The broadest bandwidth maser source known covers
$\sim$2600\kms~(Figure\,2), centered within $\sim$100\kms~of the
host galaxy systemic velocity ($V_{sys}$), and the weakest known
peak source flux densities are on the order of 10\, mJy, which
reflects the detection thresholds of existing observing systems
(Figure~1).  Because it is impossible to know {\it a priori} the
breadth or velocity of the peak maser emission from any particular
galaxy, successful observing systems combine high sensitivity and
broad bandwidth (e.g., Greenhill \etal\ 2003b).  Preponderance of
narrow-band observing systems up to 2001, posed an obstacle.
Today, the majority of survey targets have been observed with
systems capable of detecting source with peak fluxes on the order
of 10\,mJy over many thousand \kms.

Commonplace qualitative designations are: ``kilomaser,''
``megamaser,'' and ``gigamaser.'' These are used loosely in the
literature (e.g., the so-called gigamaser in TXS\,2226-182 is
apparently only $10\times$ more luminous than the common
megamaser, e.g., NGC\,3079), and classification is  based on
apparent integrated luminosity relative to that observed in local
high-mass star forming regions.

Prefixes will not be used here because of implicit ambiguity.
Specifically, maser emission is always beamed, and apparent
luminosity (L$_{\rm H_2O}$) is computed assuming isotropic
emission.  It  may exceed true luminosity by orders of magnitude.
The ratio between the two,  the beam solid angle, is determined at
least in part by source geometry and may differ widely from source
to source (e.g., a thin accretion disk beaming into a plane; a
thick fragmented disk beaming emisson into a squat cylinder; a
foreground cloud amplifies jet emission along a pencil beam).
Consequently, correlations between various quantities and apparent
maser luminosity may be anticipated to exhibit substantial scatter
(e.g., L$_{\rm H_2O}$ vs. L(2-10 keV), Kondratko \etal\ 2006b;
L$_{\rm H_2O}$ vs. X-ray absoprtion column density, Zhang \etal\
2006).  Prospectively, the discovery of any tight correlations
with L$_{\rm H_2O}$ would be notable, suggesting commonalities
among underlying physical systems. However, absent this,
qualitative (loose) correlations are still valuable, in particular
an apparent  transition at  L$_{\rm H_2O} = 1-10$\,L$_\odot$
between excitation by star formation and by nuclear activity
(Henkel \etal\ 2005).

\section{Spectroscopy}

Because VLBI follow-up is time consuming, working physical
classification of maser systems  is often made based on the
morphology of spectra.

\begin{itemize}
\item{{\bf High-velocity -- }distinct red and blue-shifted
emission complexes approximately symmetrically offset from
$V_{sys}$ ($| \rm{V-V}_{sys}|\ga200$\kms).  Perceived degrees of
symmetry vary from source to source depending on the density of
Doppler components.   Emission close to $V_{sys}$ may also be
visible, offset by $\la 10\%$ of the total source breadth. Spectra
that include high-velocity and systemic emission complexes are
distinctive indicators of emission from highly inclined accretion
disks  (Figure~2).}
\item{{\bf Low-velocity/broad -- }emission broader than
$\sim$100\kms, peaking within

\noindent $\sim$300\,km\,s$^{-1}$ of $V_{sys}$ (Figure~3).
These masers are circumstantially associated with jet activity
based on, thus far, a small sample.}
\item{{\bf Low-velocity/narrow -- }emission comprising a small
number of narrow Doppler components within $\sim$100\kms~of
V$_{sys}$.  The types of the underlying physical systems in
these cases are uncertain.}
\item{{\bf Ambiguous -- }Doppler components distributed in
velocity, including what may be high-velocity emission, but
without apparent symmetry (Figure~3).  The number of components
may be limited, which is one element that makes pattern matching
difficult.}
\end{itemize}

\begin{figure}
\begin{center}
\includegraphics[scale=0.58]{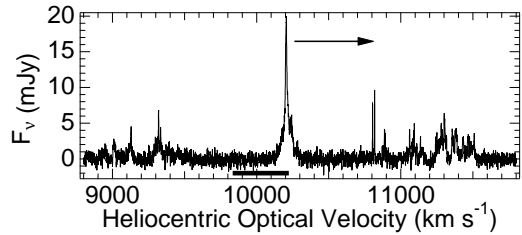}
\caption{Distinct red and blue-shifted emission complexes,
symmetry, and secular drift (arrow) in the UGC\,9618B maser. All
three are spectroscopic signatures of emission from edge-on disks.
A bar along the velocity axis indicates the range of published
V$_{sys}$ (Kondratko \etal\ 2006b).}
\end{center}
\end{figure}

\begin{figure}
\begin{minipage}{0.44\linewidth}
\centering
\includegraphics[scale=0.59]{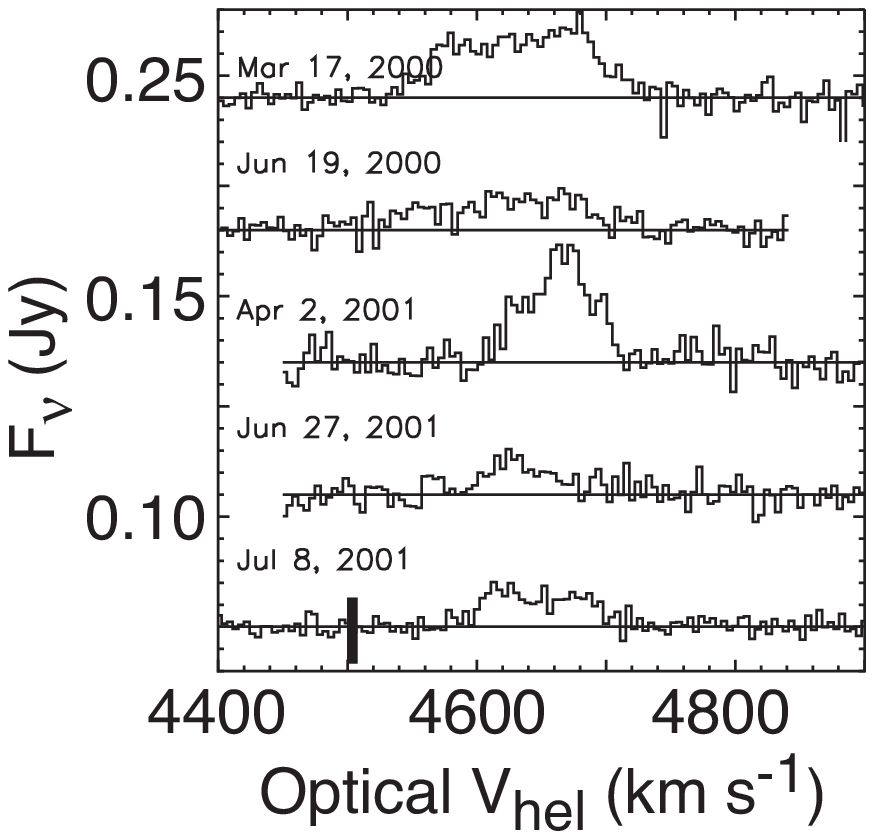}
\end{minipage}%
\begin{minipage}{0.60\linewidth}
\centering
\vspace{-0.17in}
\includegraphics[scale=0.25]{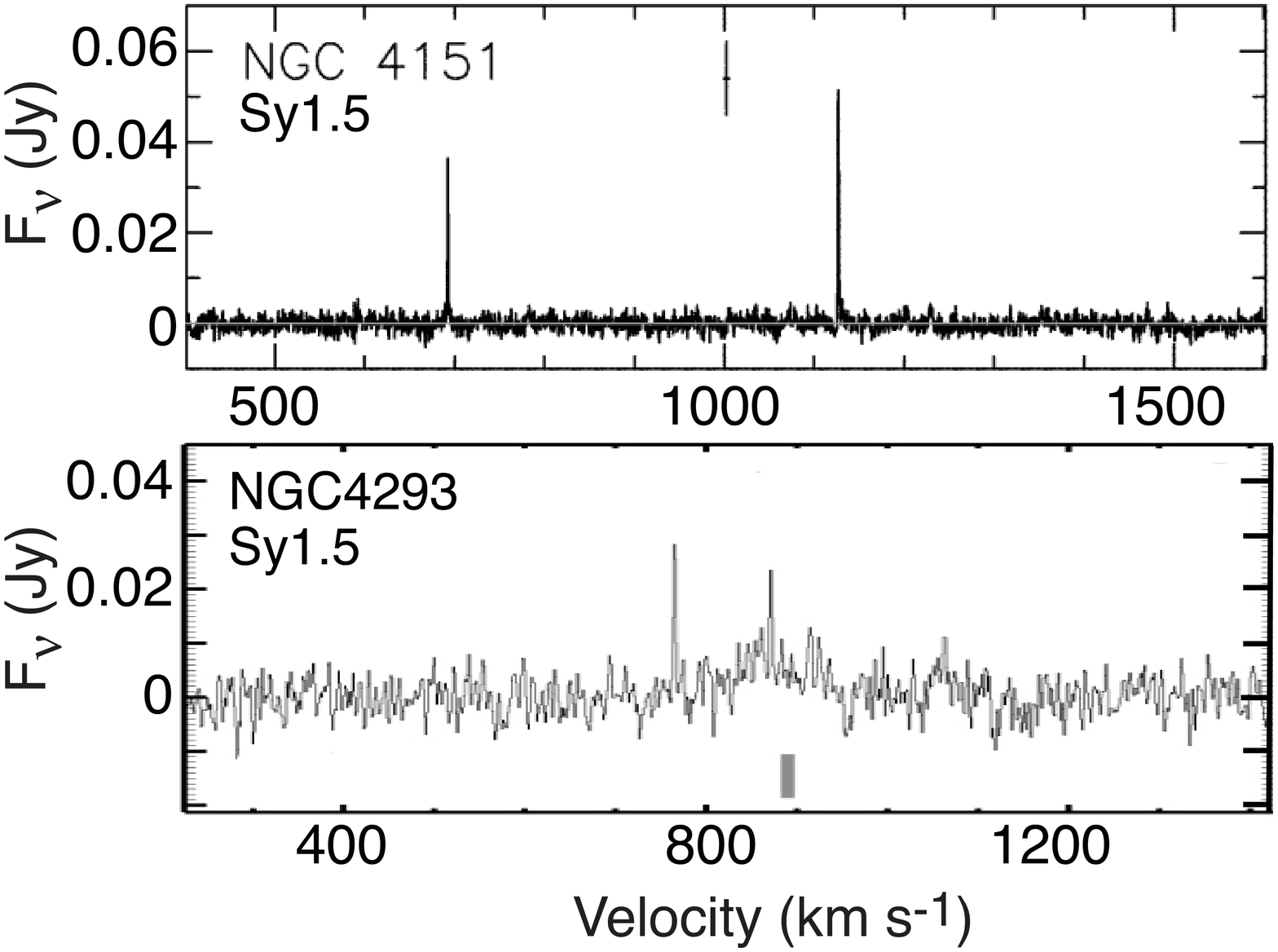}
\vspace{-0.3in}
\end{minipage}
\caption{ ({\it left})---The Mrk348 maser is associated with jet
activity and dominated by broad emission features, an empirical
signature of the origin (Peck \etal\ 2003).  ({\it right top})--A
maser with ambiguous classification; LSR/optical velocity (Braatz
\etal\ 2004).  ({\it right bottom})---a low-velocity/narrow maser;
heliocentric/radio velocity(Kondratko \etal\ 2006a). Bars mark V$_{sys}$.}
\end{figure}

All ``high-velocity'' maser systems that have been studied thus
far with very long baseline interferometry (VLBI) trace highly
inclined accretion disks or disk-structures whose rotation curves
suggest central dynamical masses of $10^6$ to $10^8$\,M$_\odot$.
These include NGC\,4258 (Miyoshi \etal\ 1995), NGC\,1068
(Greenhill \& Gwinn 1997), Circinus (Greenhill \etal\ 2003a),
NGC\,3079 (Trotter \etal\ 1998; Yamauchi \etal\ 2004; Kondratko
\etal\ 2004), NGC\,3393 (Kondratko \etal\ 2007), IC\,2560
(Ishihara \etal\ 2001; Greenhill \etal\ in prep.), and NGC\,6323
and UGC\,3789 (Braatz, this volume).  Three broad low-velocity
systems that have been mapped are offset from the central parsec
and are seen toward radio jets (NGC\,1068, Gallimore \etal\ 1996,
2004; NGC\,1052, Claussen \etal\ 1998; Mrk\,348, Peck \etal\
2003).

Supplemental classifications are also used to clarify uncertain
identifications.
\begin{itemize}
\item{{\bf Accelerating -- }long-term redward secular drift of
Doppler components close to $V_{sys}$.  This has been linked to
orbital centripetal  acceleration within inclined disks.}
\item{{\bf Symmetric -- }emission extending $\ga200$\,km\,s$^{-1}$
(symmetrically) to either side of V$_{sys}$, though perhaps
without separable emission complexes. This also marks disks.}
\item{{\bf Low-luminosity -- }apparent luminosities $\la 10$\,L$_\odot$.
Low luminosity often appears to be associated with  maser action
driven by star formation.}
\end{itemize}

For instance, lacking high-velocity emission,  the NGC\,2639 maser
classification would be ``Ambiguous'' but for observed secular
drift that suggests the maser emission originates in a disk
(Wilson \etal\ 1995).  The NGC\,1068 maser exhibits chiefly
plateau emission, but symmetry in maximum velocity offsets of $\pm
330$\,km\,s$^{-1}$ is a directy attributable to origin in a disk
(Greenhill \& Gwinn 1997).  Lastly, the M\,82 maser spectrum
exhibits just a few Doppler components near V$_{\rm sys}$, but low
apparent luminosity suggests origin in star formation (Ho \etal\
1987), as confirmed by direct imaging (Hagiwara 2007). These are
``static'' examples.  In contrast, spectra may develop defining
characteristics due to natural  source variability; an
``ambiguous'' maser with no high-velocity emission at one epoch
may display symmetric high-velocity emission at a later epoch
(e.g., NGC\,3735, Kondratko et al., in prep).  Such cases
reinforce the importance of monitoring in spectroscopic studies.

\vspace{-0.06in}
\section{Mapping}

Water maser emission  traces underlying dense, warm gas.  The
details of amplification in any individual parcel of gas may be
difficult to quantify, but nonetheless each observable maser spot
samples line-of-sight velocity on the sky, marks the sky position
of a stationary point along the line of sight in the line-of-sight
velocity field, and points to volumes where energy is deposited into
the molecular gas.  Necessarily, sampling is incomplete because maser
emission is beamed (i.e., any one observer will see only a fraction
of masers in a given source).  However, inferences from the
distribution of maser positions and velocities, coupled with {\it a priori}
constraints on geometry (e.g., largely planar, point-symmetric, rotating
structures, with circular or nearly circular orbits), enable construction
of robust models and meaningful 3D deprojections.

For a nearly edge-on, differentially rotating, well ordered, thin
($h/r \la 0.1$), flat disk that is heated by irradiation broadly
across its surface or internally via viscous dissipation, velocity
coherence favors H$_2$O maser emission: (1) in a narrow sector on
the near/far side, and (2) close to the disk-diameter
perpendicular to the line of sight (the ``midline'').  The former
gives rise to spectral features near the systemic velocity
(Figure\,2).  The latter gives rise to high-velocity emission that
traces the rotation curve of the disk when mapped with
interferometers.  NGC\,4258 is the archetype in this respect
(Herrnstein \etal\ 2005; Moran, this volume).

Emission loci for thick disks ($h/r \ga 0.1$) are comparatively
difficult to predict.  NGC\,3079 is a good example (Kondratko
\etal\ 2004).  Red and blue-shifted emission are segregated on
opposite sides of the central engine, but otherwise the maser
structure is clumped with local velocity dispersions $> 10$\% of
the rotation speed.  Between the two extremes are the relatively
well-ordered mid-range cases of Circinus, which is somewhat clumpy
but Keplerian to within modeling errors (Greenhill \etal\ 2003b),
and NGC\,1068, which is moderately clumpy and distinctly
sub-Keplerian (Greenhill \& Gwinn 1997; Lodato \& Bertin 2003).

Warps also complicate prediction of emission loci.  Maser photons
are beamed close to disk planes, (probably) with narrower beam
angles for thinner disks.  Warps broaden the solid angle into
which emission is beamed but and separate the loci where
line-of-sight velocity is coherent and where the disk is seen
edge-on.  Despite variation of $\sim$0.1\,rad in Euler angles over
a 2:1 range of radius, the  high-velocity emission loci in
NGC\,4258 are dictated by velocity coherence rather than local
tangency of the line of sight to the disk.  This could be
attributed to finite disk thickness, where in the limiting case of
a paper-thin disk, the reverse would be true.

\begin{figure}[th]
\begin{center}
\includegraphics[scale=0.45]{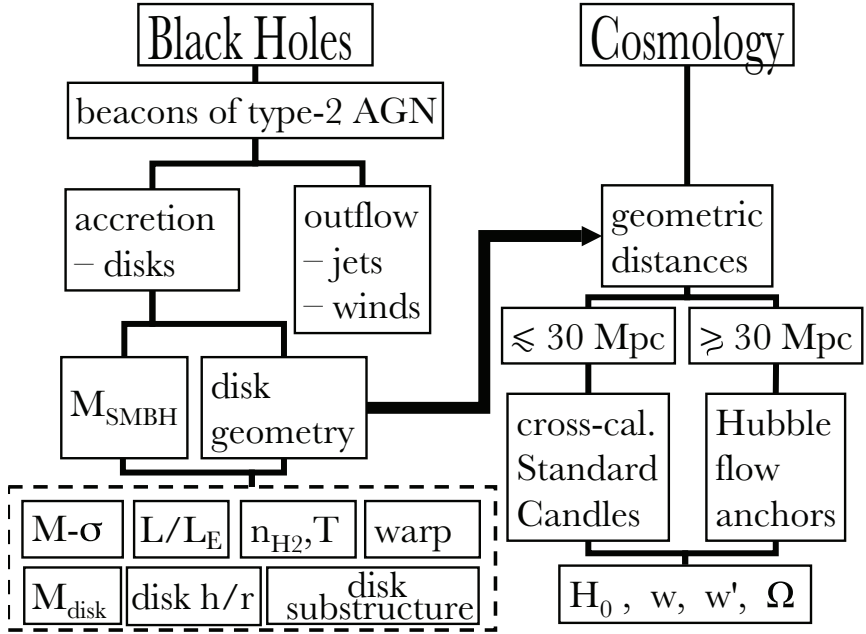}
\caption{Two threads in the science that may be accomplished through
study of H$_2$O masers  in AGN. Water masers are quite often beacons
of activity in galactic nuclei.  Related science impacts studies of
AGN physics and refinement of the extragalactic distance scale, with
ramifications for estimation of cosmological parameters.}
\label{fig:outline}
\end{center}
\end{figure}

\vspace{-0.06in}
\section{Key science}

Key science directly enabled by study of disk masers follows two
principal threads  (Figure~4):  study of AGN structure (e.g.,
Morganti \etal\ 2004) and ``maser cosmology,'' i.e.,
characterization of cosmological parameters via estimation of
accurate geometric distances to maser host galaxies (Greenhill
2004; also Braatz, this volume).  Notably, the latter depends on
the former in that  estimation of distances requires clean 3D
dynamical models of  maser disks, obtained via fitting to maser
positions and line-of-sight velocities.

{\bf AGN Structure--} Very long baseline study  of disk masers is
the only means by which structures $<1$\,pc from massive black
holes can be directly mapped.  A 1\,pc diameter annulus of maser
emission at a distance of 100\,Mpc subtends on the order of 10
interferometer beams; a maser that is 2000\kms~across can be
resolved into at least as many spectral channels with VLB
correlators.   Mapping disks in  type-2 AGN can be particularly
valuable because obscuration complicates spectroscopic and timing
analyses of central engines at other wavebands, and high
inclinations superpose reflected, scattered, and partially
obscured emission from the central engines and surrounding media
with a range of temperatures.  {\it Effectively, the AGN that are
the most difficult to study with existing instruments at other
wavebands are amenable to detailed radio study when maser emission
is present.}  The contrast is perhaps strongest for AGN with
Compton thick obscuration, for which disk masers have a strong
preference (Figure\,5).

\begin{figure}[th]
\begin{center}
\includegraphics[scale=0.60]{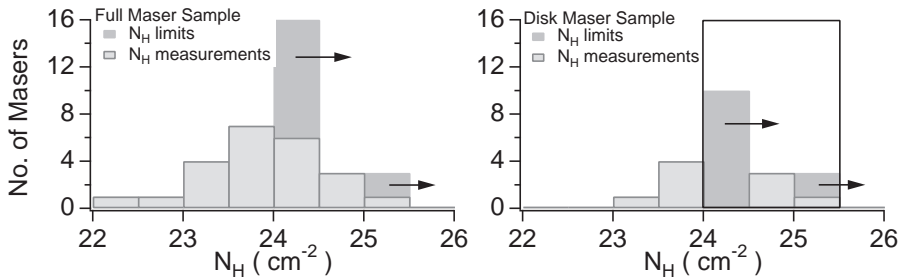}
\caption{Relatively high incidence of disk borne maser emission
among heavily obscured AGN.  {\it (left)} -- Histogram of known
X-ray absorption columns for maser AGN. {\it (right)} -- Subset
believed to originate in disks, based on evidence from spectroscopy
or mapping.  Columns are  Compton thick in 16 of 21 cases
(heavy-lined box).  See  Madejski \etal\ (2006) and Zhang \etal\ (2006).}
\end{center}
\end{figure}

Modeling VLB data in principal enables accurate measurement of
central engine dynamical masses, disk masses, Eddington
luminosities, and descriptive ratios such as  L$_{\rm
Edd}$/L$_{\rm bol}$. Dynamical masses may be used to refine
M$_\bullet-\sigma$ and M$_\bullet-M_{\rm gal}$ relations (e.g.,
Ferrarese \& Merritt 2000; Ferrarese \etal\ 2006)  that address
the question: what are the roles of massive black holes in galaxy
formation?   As well, detailed maps of disk structure may be used
to assess the balance among central engine gravity,  disk
self-gravity, luminosity, and dissipation (e.g., Tilak \etal\
2008).   Independent estimation of disk systemic velocities is
also possible, enabling refinement of galactic systemic
velocities, or estimation of central engine drift (cf. Reid \&
Brunthaler 2004). In the case of warped disks,  orientations are
known to change by 0.1-1\,rad over radii of 0.1-1\,pc.
Misalignment between central engine principle axes and reservoirs
of accreting material on parsec and larger scales can be
significant (e.g., NGC\,3079; Kondratko \etal\ 2004). Quantitative
characterization of warps enables assessment of torque origins and
disk stability  (e.g., Maloney \etal\ 1996). This also enables
assessment of warped disks as the origin of obscuring columns in
at least some,  with ramifications for the unification paradigm.
The suggestion is borne  out for NGC\,4258 (Fruscione \etal\ 2005;
Herrnstein \etal\ 2005), and circumstantial evidence for Circinus
supports the hypothesis that shadowing by warps may in some cases
bound outflows and photoionized regions (Greenhill \etal\ 2003a).

\smallskip
{\bf Cosmology--} The Standard Model is flat and comprises
radiation, baryonic matter, nonbaryonic matter, and dark energy.
It has been  defined largely by analyses of Cosmic Microwave
Background fluctuations  and supernova distances (e.g., Spergel
\etal\ 2007).  Nevertheless, estimates of geometry and the
physical nature of dark energy remain too uncertain to distinguish
among contending theories without substantive assumptions;
external constraints are required, as from S-Z effect or lensing.
High-accuracy measurement of the Hubble constant, $H_\circ$, would
be another and arguably a more direct one. Accuracy of $\sim$1\%
would enable strong constraint on the dark energy equation of
state (Hu 2005; Macri \etal\ 2006).  The present best independent
estimate of $H_\circ$ is perhaps accurate to 10\%, though
systematic errors are difficult to quantify (Argon \etal\ 2007).

A sample of a few to a few tens of anchor galaxies with measured
geometric distances could support direct estimation of H$_\circ$,
probably with higher accuracy than is possible with traditional
Standard Candles, considering intrinsic and systematic
uncertainties.   All distance estimates for $N$ anchors  being
equally good, uncertainties in $H_\circ$ would scale as $N^{-0.5}$
because systematic uncertainties for disk models among galaxies
depend chiefly on individual emission distributions and are
uncorrelated.   In fact, focus on a few very good cases, e.g.,
NGC\,4258-like galaxies in the Hubble flow, might yield better
results overall than a larger sample of more middling cases,
though the  distribution in $cz$ matters.

In principal, two geometric distance estimates may be obtained for
each maser galaxy, one via analysis of centripetal acceleration,
and one via analysis of proper motion (Herrnstein \etal\ 1999).
Both observables are greatest for masers close to the systemic
velocity.  Measurement of acceleration is easier and less resource
intensive.  (Proper motions are a very small fraction of a
beamwidth for earth-bound baselines; 1000\kms\ subtends
$\sim$$2\mu$as\,yr$^{-1}$ at $\sim$100\,Mpc, or $\sim$1\% of a
beam at $\lambda1$\,cm.   For an orbital radius of 1\,pc, this
speed corresponds to a $\sim$1\,km\,s$^{-1}$\,yr$^{-1}$ drift or
about one line width per year.)  Nonetheless, VLB study is
requisite to establish a disk rotation curve and model from which
to estimate central mass divided by distance and deprojected
orbital radii for systemic emission.  The latter are distinct from
the radii of high-velocity emission inferred from rotation curves
and in first order analyses may be inferred from the rate of
change of line-of-sight velocity as a function of position along
the disk plane.

The basic methodology to measure the geometric distance to an
anchor galaxy involves several VLB epochs, coupled with
single-dish monitoring of spectra over a time interval that
includes the VLB observations (to enable cross-referencing).
Considering time variability of Doppler components over weeks to
months, VLB epochs need to be closely spaced in time (in the event
more than one is required to achieve adequate sensitivity).
However, widely spaced clusters of VLB epochs are also useful in
cases where the velocity pattern of Doppler components evolves
(i.e., new lines appearing at new velocities).  Repeated VLB
observation may broaden the fraction of the disk that can be
traced  (e.g., by sampling the midline over a larger range of
radii), thus improving model quality.

VLB modeling efforts thus far suggest the best and highest
priority masers for estimation of extragalactic distances would be
those that conform to five basic criteria:
\begin{itemize}
\item{red {\it and} blue-shifted high-velocity emission --}
\subitem{\bp many high-velocity lines sampling rotation curves over a
broad range of radius; }
\item{emission close to the systemic  velocity --}
\subitem{\bp many lines near systemic that display measurable
centripetal accelerations;}
\subitem{\bp measurable position-velocity gradients from which
radii may be inferred;}
\item{thin disk structure --}
\subitem{\bp emission divisible by eye into narrow Doppler components;}
\subitem{\bp Keplerian rotation (i.e.,  little evidence for
fragmentation or other perturbation);}
\subitem{\bp a rich emission  distribution that enables robust
modeling, including independent}
\subitem{~~~ estimation of disk inclination, systemic velocity,
and dynamical center location; }
\item{peculiar motion that is small with respect to total
recessional velocity --}
\subitem{\bp origin in the Hubble Flow;}
\item{a line strong enough for VLB self-calibration {\it or}  a
strong calibrator $\ll 1^\circ$ away.}
\end{itemize}

\end{document}